# Drude weight and dc-conductivity of correlated electrons

G. Uhrig[a] and D. Vollhardt[b]

[a]*Institut für Theoretische Physik, Universität zu Köln D-50937 Köln, Germany*

[b]*Institut für Theoretische Physik C, RWTH Aachen, D-52056 Aachen, Germany*

(April 5, 1995)

## Abstract

The Drude weight $D$ and the dc-conductivity $\sigma_{dc}(T)$ of strongly correlated electrons are investigated theoretically. Analytic results are derived for the homogeneous phase of the Hubbard model in $d = \infty$ dimensions, and for spinless fermions in this limit with $1/d$-corrections systematically included to lowest order. It is found that $\sigma_{dc}(T)$ is finite for all $T > 0$, displaying Fermi liquid behavior, $\sigma_{dc} \propto 1/T^2$, at low temperatures. The validity of this result for finite dimensions is examined by investigating the importance of Umklapp scattering processes and vertex corrections. A finite dc-conductivity for $T > 0$ is argued to be a generic feature of correlated lattice electrons in not too low dimensions.

# I. INTRODUCTION

Strongly correlated electron systems, such as heavy fermions and high-$T_c$ superconductors, have very unusual thermodynamic and transport properties. To understand the physical mechanisms underlying these properties, the temperature-dependent dynamic ("optical") conductivity $\sigma(\omega, T)$ is one of the most meaningful physical quantities to be studied both experimentally and theoretically. Of special interest is the low-frequency behavior of the conductivity since it provides information about scattering processes close to the Fermi level. Theoretical studies of $\sigma(\omega, T)$ for $\omega \to 0$ so far concentrated on the limit $T = 0$ [1–8] where the real part of the conductivity can be written as

$$\lim_{T \to 0} \mathrm{Re}\ \sigma(\omega, T) = D\delta(\omega) + \sigma_{\mathrm{incoh}}(\omega)\ . \qquad (1)$$

Here $D$ is the weight of the Drude peak and $\sigma_{\mathrm{incoh}}$ is an incoherent background. As pointed out by Kohn [1] the Drude weight $D$ is of particular importance since it provides a direct and sensitive criterion for a metal-insulator transition.

The behavior of the dc-conductivity

$$\sigma_{dc}(T) = \sigma(\omega = 0, T) \qquad (2)$$

at $T > 0$ is not so well understood theoretically and received attention only most recently [9–13]. If a system has *continuous* translational invariance $\sigma_{\mathrm{dc}}$ must be infinite even at finite temperatures since the momentum acquired from the external electric field cannot be degraded (here we do not consider scattering with phonons or impurities). Indeed, the Hamiltonian $\hat{H}$ commutes with the total momentum $\hat{\vec{P}}$ operator

$$[\hat{H}, \hat{\vec{P}}] = 0. \qquad (3)$$

Here and in the following operators carry a hat. Hence any expectation value of $\hat{\vec{P}}$ and thus of the current operator $\hat{\vec{J}} = -e\hat{\vec{P}}/(mL_1)$, is constant in time (here the electron charge is $-e$, $m$ is the electron mass and $L_1$ is the length of the sample in the direction of $\hat{\vec{P}}$). No damping occurs.

For a *lattice* system, characterized by a tight-binding Hamiltonian, this argument no longer holds. The current operator and the Hamiltonian do not commute and there is no reason why the conductivity should not be finite. Yet even on a lattice there are constraints on the destruction of momentum: the crystal momentum is conserved modulo reciprocal lattice vectors. In the absence of phonons or impurities crystal momentum can only be destroyed by *Umklapp* scattering. Now, it is not a priori clear whether Umklapp processes alone are sufficient to reduce $\sigma_{dc}(T)$ to a finite value. For one-dimensional fermion systems this problem has been investigated in detail by Giamarchi [9]. In this paper we will approach the question of whether or not the dc-conductivity of correlated electrons may diverge at $T > 0$ from the opposite direction, i.e. the limit of *high* dimensions.

In the following we consider a general lattice Hamiltonian of the form

$$\hat{H} = \sum_{i,j;\sigma} \frac{t_{ij}}{\sqrt{Z_{ij}}} \hat{c}^+_{i\sigma} \hat{c}_{j\sigma} + \frac{1}{2} \sum_{ij} \frac{U_{ij}}{Z_{ij}} \hat{n}_i \hat{n}_j \qquad (4)$$

where $\hat{c}^+_{i\sigma}$ ($\hat{c}_{i\sigma}$) is the creation (annihilation) operator of a electron with spin $\sigma$ at site $i$ and $\hat{n}_i = \sum_\sigma \hat{c}^+_{i\sigma} \hat{c}_{i\sigma}$ is the total particle density at site $i$. The coefficients $t_{ij}$ and $U_{ij}$ are hopping and interaction matrix elements, respectively. The coordination numbers $Z_{ij}$ count the number of sites which are in the same point-symmetry group as $j$ with respect to site $i$. We assume the system to be invariant under discrete lattice translations and transformations of the point-symmetry groups of each site. In all explicit calculations we restrict ourselves to nearest-neighbor hopping on a hypercubic lattice (with coordination number $Z = 2d$, where $d$ is the dimension), i.e. assume $t_{ij} \equiv -t^*$ for nearest-neighbors and $t_{ij} = 0$ otherwise. We set $t^* = 1$, i.e. measure all energies in units of $t^*$. The kinetic energy then reads

$$\hat{H}_{kin} = -\frac{1}{\sqrt{Z}} \sum_{\langle ij \rangle, \sigma} \hat{c}^+_{i\sigma} \hat{c}_{j\sigma} = \sum_{\vec{k}, \sigma} \epsilon_{\vec{k}} \hat{n}_{\vec{k}\sigma} \qquad (5a)$$

where $\epsilon_{\vec{k}}$ is the energy dispersion,

$$\epsilon_{\vec{k}} = -\frac{2}{\sqrt{Z}} \sum_{i=1}^d \cos k_i \qquad (5b)$$

with unit lattice spacing, and $\hat{n}_{\vec{k}\sigma}$ denotes the moment distribution operator. Due to particle-hole symmetry it is sufficient to discuss densities $n = N/L$ not larger than half-filling. Here $N$ is the total number of particles and $L$ is the number of lattice sites. The doping $\delta$ is measured relative to half-filling, i.e. $\delta = 1 - n$ for the Hubbard model and $\delta = \frac{1}{2} - n$ for spinless fermions.

The scaling of the hopping by the square root of the coordination numbers in (4) is essential to obtain a well-defined limit $d \to \infty$ for fermionic lattice models [14]. The scaling of the interaction terms in (4) is more straightforward: it must be chosen such that the total interaction strength over all sites does not diverge [15]. For explicit calculations we will consider electrons (with spin $\sigma = \uparrow, \downarrow$) interacting via a Hubbard interaction

$$\hat{H}_{Hubbard} = -\frac{1}{\sqrt{Z}} \sum_{\langle ij \rangle, \sigma} \hat{c}^+_{i\sigma} \hat{c}_{j\sigma} + U \sum_i \hat{n}_{i\uparrow} \hat{n}_{i\downarrow}, \qquad (6a)$$

as well as spinless fermions (no spin index) with nearest-neighbor interaction

$$\hat{H}_{spinless} = -\frac{1}{\sqrt{Z}} \sum_{\langle ij \rangle} \hat{c}^+_i \hat{c}_j + \frac{U}{2Z} \sum_{\langle ij \rangle} \hat{n}_i \hat{n}_j . \qquad (6b)$$

The limit $d \to \infty$ is carried out on the level of ordinary diagrammatic perturbation theory where it leads to considerable simplifications [16]. These simplifications make an analytic solution of (6b) possible since the Hartree approximation becomes exact in this case [15,17–19].

By contrast, the Hubbard model remains dynamic even in $d = \infty$ [15] and thus cannot be solved analytically. It reduces to a single-site theory [22] which is equivalent to an effective single-impurity problem with a self-consistency condition [23,24]. This equivalence makes numerical [24–27] and approximate [23,27] solutions possible.

In the following we will investigate the dc-conductivity $\sigma_{dc}(T)$ for the two models (6a,b) in the limit of large dimensions $d$, i.e. large coordination number $Z$ (this limit is denoted by $d \to \infty$ or, equivalently, by $Z \to \infty$). In the case of the Hubbard model we work strictly at $Z = \infty$. However, in the case of spinless fermions all terms up to order $1/Z$ are explicitly included. Not only is this managable in this case [19], but - more importantly - it is mandatory for obtaining finite results for $\sigma$: in the limit $Z = \infty$ the theory is purely static (Hartree theory), leading to an infinite conductivity, since scattering processes are absent. The latter first enter into the theory through the inclusion of $1/Z$-corrections [28], whereby genuine correlation terms are generated. We will refer to this limit as "spinless fermions with $1/Z$-terms".

To be able to study the effects of correlations on $\sigma_{dc}(T)$ at low temperatures we will deliberately ignore the onset of long-range order (i.e. spin and charge order in the Hubbard model and spinless fermion model, respectively), since the existence of an energy gap would naturally lead to an exponential suppression of $\sigma_{dc}(T)$ for $T \to 0$.

The paper is structured as follows: in Section II the dynamic conductivity $\sigma(\omega, T)$ is calculated in the limit of high dimensions. A detailed investigation of the limit $\omega \to 0$, i.e. of the Drude weight $D$ at $T = 0$ and the dc-conductivity $\sigma_{dc}(T)$ at $T > 0$, is presented in Section III. Explicit results are obtained for the Hubbard model in $d = \infty$ and spinless fermions with $1/Z$-terms. The finiteness of $\sigma_{dc}(T)$ at $T > 0$ and its validity for dimensions $d < \infty$ is discussed in Section IV by examining the importance of Umklapp scattering processes and vertex corrections.

## II. CONDUCTIVITY $\sigma(\omega, T)$ IN THE LARGE-$D$ LIMIT

The dynamic conductivity can be calculated from the Kubo formula as

$$\sigma(\omega, T) = g\Big[\sigma_1(\omega, T) + \sigma_2(\omega, T)\Big] \tag{7a}$$

where

$$\sigma_1(\omega, T) = \frac{ie^2}{\omega} \int_{BZ} \frac{d^d k}{(2\pi)^d} \frac{\partial^2 \varepsilon_{\vec{k}}}{\partial k_1^2} \langle \hat{n}_{\vec{k}} \rangle \tag{7b}$$

$$\sigma_2(\omega, T) = \frac{ie^2}{\omega} \chi^{JJ}(\omega, T). \tag{7c}$$

Here $g$ is a spin degeneracy factor, i.e. $g = 1$ for spinless fermions and $g = 2$ for the Hubbard model and $\chi^{JJ}(\omega)$ is the retarded current-current correlation function; the current is chosen to flow in the 1-direction. Since the current flows into a given direction, i.e. is proportional to the effective hopping matrix-element which itself is of order $1/\sqrt{Z}$ small, the current-current correlation function, and thereby the conductivity $\sigma$ itself, is proportional to the square of the effective hopping. Hence it is of order $1/Z$ small. Strictly speaking the scaling of the hopping would therefore imply that the conductivity – like any other transport coefficient – would vanish in the limit $Z \to \infty$. In the following we therefore investigate the leading order contribution in $1/Z$ to $\sigma$, i.e. $\lim_{Z \to \infty} Z\sigma(\omega, T)$ which essentially represents the sum of that quantity over an entire shell of nearest-neighbor sites.

For the infinite dimensional Hubbard model the conductivity is given diagrammatically by the dressed bubble-diagram [29,30,10,18]. Vertex corrections do not enter since they are of order $1/Z^2$. The conductivity is hence given by a convolution of two one-particle propagators $G_{\vec{k}}(i\omega_n) \equiv G_{\vec{k},n} = [i\omega_n + \mu - \Sigma_n - \epsilon_{\vec{k}}]^{-1}$ where $\Sigma_n \equiv \Sigma(i\omega_n)$ is the self-energy (which is $\vec{k}$-independent in $d = \infty$ [14,15]) and $\mu$ is the chemical potential, i.e.

$$\chi^{\text{JJ}}(i\omega_n, T) = \frac{4T}{Z} \sum_{m,l} \int_{\text{BZ}} \frac{d^d k}{(2\pi)^d} \sin^2(k_1) G_{\vec{k},m} G_{\vec{k},l} \delta_{\omega_n, \omega_m - \omega_l} . \tag{8a}$$

For spinless fermions with $1/Z$-terms the r.h.s. of (8a) yields the auxiliary function $\chi_0^{\text{JJ}}(i\omega_n, T)$ from which the full correlation function follows as [28]

$$\chi_{spinless}^{\text{JJ}}(i\omega_n, T) = \frac{2\chi_0^{\text{JJ}}(i\omega_n, T)}{2 + U\chi_0^{\text{JJ}}(i\omega_n, T)} . \tag{8b}$$

In the limit $d \to \infty$ the term $\sin^2(k_1)$ can be replaced by its average in the Brillouin zone, i.e. by $1/2$. Then the integrand in (8a) depends on $\vec{k}$ only through the energy dispersion $\epsilon_{\vec{k}}$ of the $non$-interacting electrons. This demonstrates clearly the mean-field character of the $d = \infty$ limit. The density of states (DOS) of the non-interacting system, $N_0(\epsilon)$, which for a hypercubic lattice (see (5b)) in $d = \infty$ takes the form [14]

$$N_0^\infty(\varepsilon) = \frac{1}{\sqrt{2\pi}} \exp(-\varepsilon^2/2), \tag{9}$$

may then be used to express (8a) as an energy integral

$$\chi^{\text{JJ}}(i\omega_n, T) = \frac{2T}{Z} \sum_{m,l} \int_{-\infty}^{\infty} d\epsilon N_0(\varepsilon) G_m(\varepsilon) G_l(\varepsilon) \delta_{\omega_n, \omega_m - \omega_l} , \tag{10}$$

where $G_n(\varepsilon) = [i\omega_n + \mu - \Sigma_n - \varepsilon]^{-1}$. If the self-energy $\Sigma_n$ is known the best way to proceed is to use a partial fraction expansion

$$G_m(\varepsilon) G_l(\varepsilon) = -[(i\omega_m - \Sigma_m) - (i\omega_l - \Sigma_l)]^{-1} [G_m(\varepsilon) - G_l(\varepsilon)] \tag{11}$$

which permits one to perform the $\varepsilon$-integral in (10) as

$$\chi^{\mathsf{JJ}}(i\omega_n, T) = -\frac{2T}{Z} \sum_{m,l} \frac{g_m - g_l}{i\omega_m - i\omega_l - (\Sigma_m - \Sigma_l)} \delta_{\omega_n, \omega_m - \omega_l} \tag{12}$$

where $g_m = \int d\varepsilon N_0(\varepsilon) G_m(\varepsilon)$ is the local one-particle propagator. Analytic continuation then leads to

$$\chi^{\mathsf{JJ}}(\omega, T) = \frac{1}{iZ\pi}(B_1 + B_2), \tag{13}$$

where [28]

$$B_1(\omega, T) = \int_{-\infty}^{\infty} d\omega' [f(\omega' + \omega) - f(\omega')] \frac{g_R(\omega' + \omega) - g_A(\omega')}{\omega - [\Sigma_R(\omega' + \omega) - \Sigma_A(\omega')]} \tag{14a}$$

$$B_2(\omega, T) = \int_{-\infty}^{\infty} d\omega' f(\omega') \left\{ \frac{g_R(\omega' + \omega) - g_R(\omega')}{\omega - [\Sigma_R(\omega' + \omega) - \Sigma_R(\omega')]} - \frac{g_A(\omega') - g_A(\omega' - \omega)}{\omega - [\Sigma_A(\omega') - \Sigma_A(\omega' - \omega)]} \right\}. \tag{14b}$$

Here $g_R$ and $g_A$ are the retarded and advanced local one-particle propagators, respectively, and $f(x) = (e^{\beta x} + 1)^{-1}$ is the Fermi function. In the limit $T, \omega \to 0$ the term $B_1$ dominates since it reduces to a constant and hence leads to a divergent contribution to $\sigma_2(\omega, T)$, (7c), and thereby to $\sigma(\omega, T)$ itself. However, in the case of spinless fermions with $1/Z$-terms $B_2$ has to be taken into account, too (see below).

### III. THE LIMIT $\omega \to 0$

#### A. $T = 0$: study of the Drude weight $D$

To obtain an analytic expression for the Drude weight $D$ in (1) we use (14a) for $B_1$ in the limit of small $\omega$

$$B_1(\omega, 0) = 2\pi i N(0) \frac{\omega}{(m^*/m)\omega + i\eta}. \tag{15}$$

Here $N(0)$, the value of the DOS $N(\omega) = -\pi^{-1} \mathrm{Im} g_R$ of the interacting electrons at the Fermi level, enters. Furthermore $m^*$, with $m^*/m = 1 - d\mathrm{Re}\Sigma/d\omega|_{\omega=0}$, is the effective mass, and $i\eta$ is an imaginary part which is infinitesimally small at $T = 0$ but becomes finite at finite temperatures. From (7a) we find for $D$

$$D = 2 \int_{-\eta'}^{\eta'} d\omega \, \mathrm{Re}\sigma_2(\omega, 0) \tag{16}$$

where $\eta'$ is also infinitesimally small at $T = 0$. The contribution of $\sigma_1$, (7b), need not be considered here since it is purely imaginary. For the same reason $B_2$ is unimportant in the evaluation of $D$ for the Hubbard model. Combination of (15) and (16) yields

$$ZD_{Hubbard} = \frac{4\pi e^2}{m^*/m} N(0). \tag{17}$$

The situation is different in the case of spinless fermions with $1/Z$-terms [28]. Here the limiting value of $B_2 = \pi i \gamma^{-1} \langle \hat{H}_{kin} \rangle$ for small $\omega$ must be taken into account, too, where

$$\gamma = 1 - \sqrt{Z} \Sigma^{Fock} \tag{18}$$

is a renormalization factor resulting from the frequency-independent Fock-contribution to the self energy [31]. Furthermore, $N(\omega)$ must be replaced by the spectral density

$$N_c(\omega) = -\frac{1}{\pi} \text{Im } c(\omega) \tag{19a}$$

defined through

$$c(\omega) = \int_{-\infty}^{\infty} d\varepsilon \frac{N_{c,0}(\varepsilon)}{\omega + \mu - \Sigma(\omega) - \varepsilon} \tag{19b}$$

with $N_{c,0}(\varepsilon)$ determined from $N_0(\omega)$ as

$$\frac{dN_{c,0}}{d\omega} = \frac{\omega}{2} N_0(\omega). \tag{19c}$$

In $d = \infty$, with $N_0(\omega)$ given by (9), this implies $N_{c,0}^\infty(\varepsilon) = \frac{1}{2} N_0^\infty(\varepsilon)$ and therefore $c_{R,A}(\omega) = \frac{1}{2} g_{R,A}(\omega)$ and $N_c(\varepsilon) = \frac{1}{2} N(\varepsilon)$. Finally, by increasing the effective hopping of spinless fermions with $1/Z$-terms [31] the factor $\gamma$ also *reduces* the effective mass, i.e. $m^*/m = (1 - d\text{Re}\Sigma/d\omega|_{\omega=0})/\gamma$. In this way one finds in analogy to (17)

$$ZD_{spinless} = \frac{4\pi e^2 \gamma N_c(0)}{m^*/m + 2Z^{-1} U N_c(0)}. \tag{20}$$

For the evaluation of (17) and (20) at $T = 0$ it is helpful to note that in $d = \infty$ the quantities $\mu - \Sigma(0)$ (for the Hubbard model) and $[\mu - \Sigma(0)]/\gamma$ (for spinless fermions with $1/Z$-terms) are independent of $U$. This is a direct consequence of Luttinger's theorem on the constancy of the volume within the Fermi surface [32] and the fact that $\Sigma$ is strictly local (Hubbard model). It even holds when $\Sigma$ contains terms which are *proportional* to the kinetic energy (as in the case of $\Sigma^{Fock}$ for spinless fermions with $1/Z$-terms), even when they are frequency dependent as in the case of the Hubbard model with $1/Z$ corrections. Consequently, $N(0)$ in (17) and $\gamma N_c(0)$ in (20) may be replaced by $N_0(0)$ and $N_{c,0}(0)$, respectively, i.e. by the quantities at $U = 0$, at the same filling.

The scaled Drude weight ZD for spinless fermions with $1/Z$-terms, obtained by evaluating the r.h.s. of (20) in $d = 3$ (i.e. $Z = 6$) with the actual three-dimensional DOS $N_0(\varepsilon)$ and $N_{c,0}(\varepsilon)$, is shown in Figs. 1,2 as a function of interaction $U$ and doping $\delta$, respectively. Also included in Fig. 1 is the total weight $Z \int d\omega \sigma(\omega)$ at $\delta = 0$ which is essentially equal to the kinetic energy $Z \int_{-\infty}^{\infty} d\omega \sigma(\omega) = -2\pi e^2 \langle \widehat{H}_{kin} \rangle$ ($f$-sum rule) [33,34]. The difference between the two curves for $\delta = 0$ represents the weight outside the $\delta$-peak, i.e. $\int_{-\infty}^{\infty} d\omega \sigma_{\text{incoh}}(\omega)$. Both ZD and the total weight decrease quadratically with $U$ at small $U$ (Fig. 1). A large $U$, however, $ZD \propto (\ln U)^{-1}$ which implies a very slow decrease. The latter behavior is a consequence of the suppression of long-range order: in the symmetry-unbroken phase $\Sigma^{Fock}$ *adds* to the hopping such that extended states are favored even at large $U$.

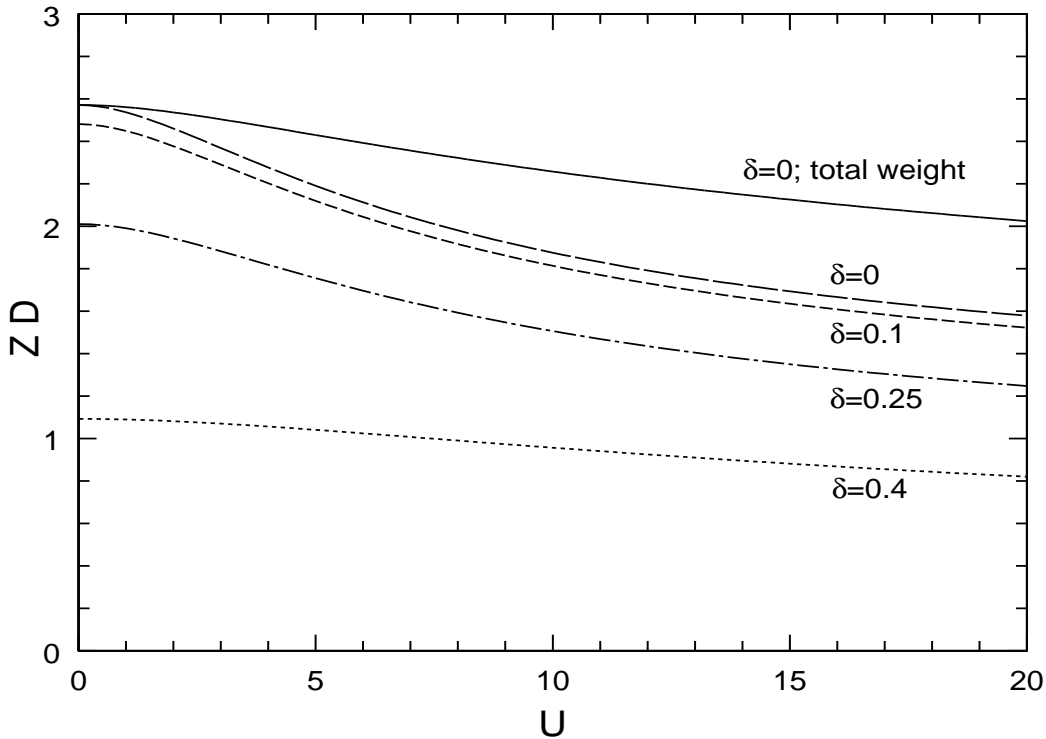

FIG. 1. Scaled Drude weight $ZD$ vs. interaction $U$ for spinless fermions with $1/Z$-terms at $T=0$ for different fillings ($D$ in units of $e^2 t^*/(\hbar a^{d-2})$, where $a$ is the lattice constant; $U$ in units of $t^*$). Solid curve: total weight $\int d\omega \sigma(\omega)$ (same units as $D$) for $\delta = 0$.

While the decrease of the Drude weight with increasing repulsion is to be expected, its $\delta$-dependence – namely the global suppression of $ZD$ upon doping – needs explanation. The latter effect is seen most clearly in Fig. 2 where $ZD$ is found to decrease monotonically as a function of $\delta$. This is surprising since close to half filling (small $\delta$) one would expect doping to *improve* the mobility of the particles, as it is observed in the case of the Hubbard model [2,4,6]. Here the essential difference between the interaction in the Hubbard model and in the spinless fermion model comes into play. Let us consider a half-filled band at not too small $U$. In the Hubbard model the lattice sites are then singly occupied. The particles are thereby essentially *localized*, i.e. the Drude weight $D$ vanishes. This effect is independent of whether the spins are antiferromagnetically ordered or not. Spin ordering, through which the energy is lowered by a (small) amount $\sim t^2/U$, is here an *additional* effect with rather little influence on $D$, since $D$ is already essentially zero. By contrast, for spinless fermions such a localization effect, and hence $ZD \simeq 0$ at $\delta = 0$, can *only* occur in the presence of long-range charge-order (checker board structure on the two sublattices) since then the nearest-neighbor interaction is effectively avoided. In this case doping would indeed improve the mobility and thus increase $ZD$. In the homogeneous phase (which we consider here precisely because we wish to ignore the *obvious* effects of long-range order; see the discussion at the end of Section I) this localization effect is absent and hence $ZD$ always decreases as a function of $\delta$. At $\delta = 0.29$ there is a weak cusp due to the van Hove singularity of the DOS in $d = 3$ at one third of the band width. Note that in the empty band limit ($\delta \to 0.5$) $ZD$ vanishes with the same slope for all $U$ since the effect of the interaction decreases with decreasing particle density.

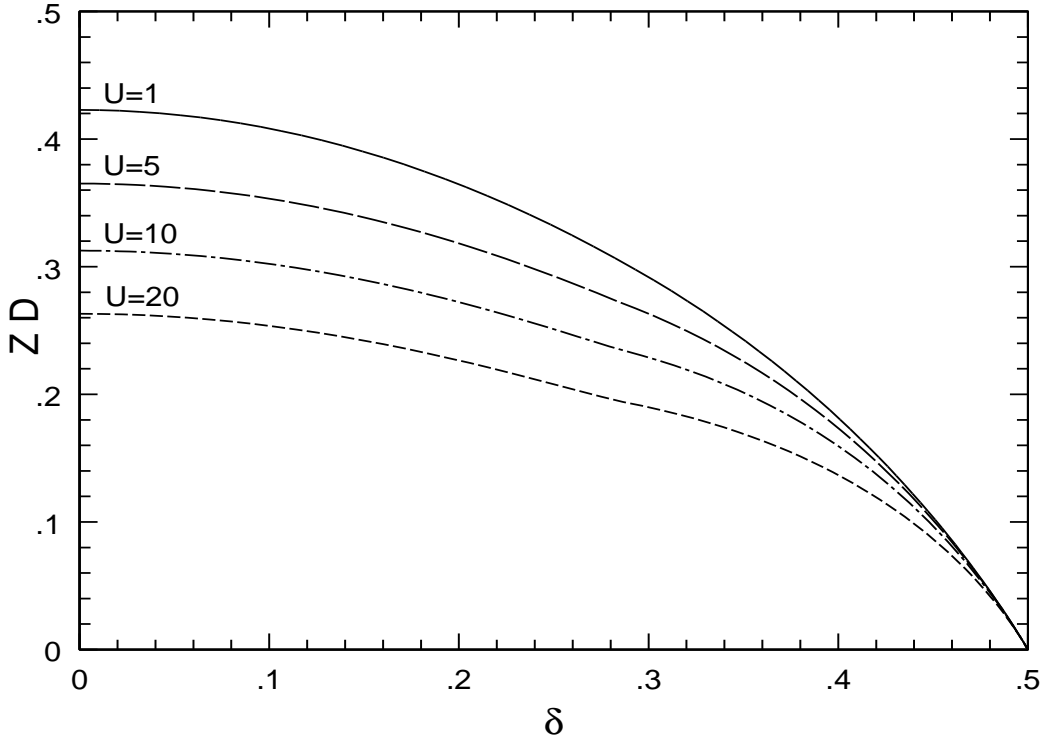

FIG. 2. Scale Drude weight $ZD$ vs. doping $\delta$ for spinless fermions with $1/Z$-terms at $T = 0$ for different $U$ values; same units as in Fig. 1.

### B. $T > 0$: study of the dc-conductivity

At low temperatures the dc-conductivity for the Hubbard model is found from (7), (13), (14) as

$$Z\sigma_{dc}^{Hubbard}(T) = \frac{e^2}{\pi} \int_{-\infty}^{\infty} d\omega \frac{N(\omega)}{N_\Sigma(\omega)} \left( -\frac{\partial f(\omega)}{\partial \omega} \right), \qquad (21)$$

where $N_\Sigma(\omega) = -\pi^{-1}\mathrm{Im}\Sigma_R(\omega)$.

In analogy, for spinless fermions with $1/Z$-terms one obtains

$$Z\sigma_{dc}^{spinless}(T) = \frac{e^2\gamma^2}{\pi} \int_{-\infty}^{\infty} d\omega \frac{N_c(\omega)}{N_\Sigma(\omega)} \Big( -\frac{\partial f(\omega)}{\partial \omega} \Big) . \quad (22)$$

Here we encounter a peculiarity due to the fact that in the case of spinless fermions the limits $Z \to \infty$ and $\omega \to 0$ do not commute: although in the limit $Z \to \infty$ the dynamic conductivity $Z\sigma^{spinless}(\omega, T) \sim \mathcal{O}(1)$ for all $\omega \neq 0$, and also $Z \int_{-\infty}^{\infty} d\omega \sigma^{spinless}(\omega, T) \sim \mathcal{O}(1)$, the dc-conductivity behaves differently since $N_\Sigma(\omega) \sim \mathcal{O}(1/Z)$. Equ. (22) then implies that $Z\sigma_{dc}^{spinless}(T) \sim \mathcal{O}(Z)$ for $Z \to \infty$, i.e. the resistivity obeys $\rho^{spinless}(T)/Z \sim (1/Z)$, while $\rho^{Hubbard}(T)/Z \sim \mathcal{O}(1)$.

The results for $\rho(T)/Z$ in the case of spinless fermions with $1/Z$-terms at half filling, obtained by evaluating the inverse of the r.h.s. of (22) in $d = 3$ ($Z = 6$), are shown in Figs. 3 and 4. They are compared with the results of Pruschke, Cox and Jarrell [10] for the half-filled Hubbard model in $d = \infty$, obtained in the non-crossing approximation. In both cases $U = 4.243$. (We note that for the Hubbard model this value of $U$ is just below the value where the Mott-Hubbard transition occurs [25,26,10]; the system is therefore quite close to the transition).

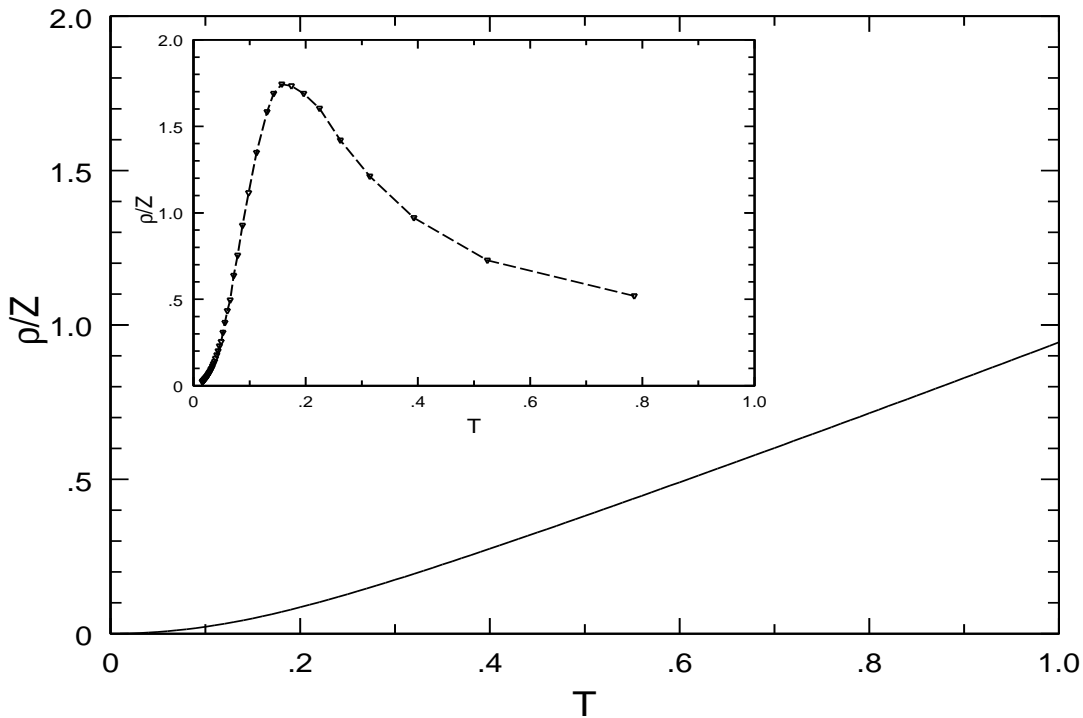

FIG. 3. Scaled resistivity $\rho/Z$ vs. temperature $T$ at $U = 4.243t^*$ and $\delta = 0$ ($\rho$ in units of $\hbar a^{d-2}/e^2$; $T$ in units of $t^*$). Main figure: spinless fermions with $1/Z$-terms in $d = 3$. Inset: Hubbard model in $d = \infty$, evaluated in the non-crossing approximation [8]; see text.

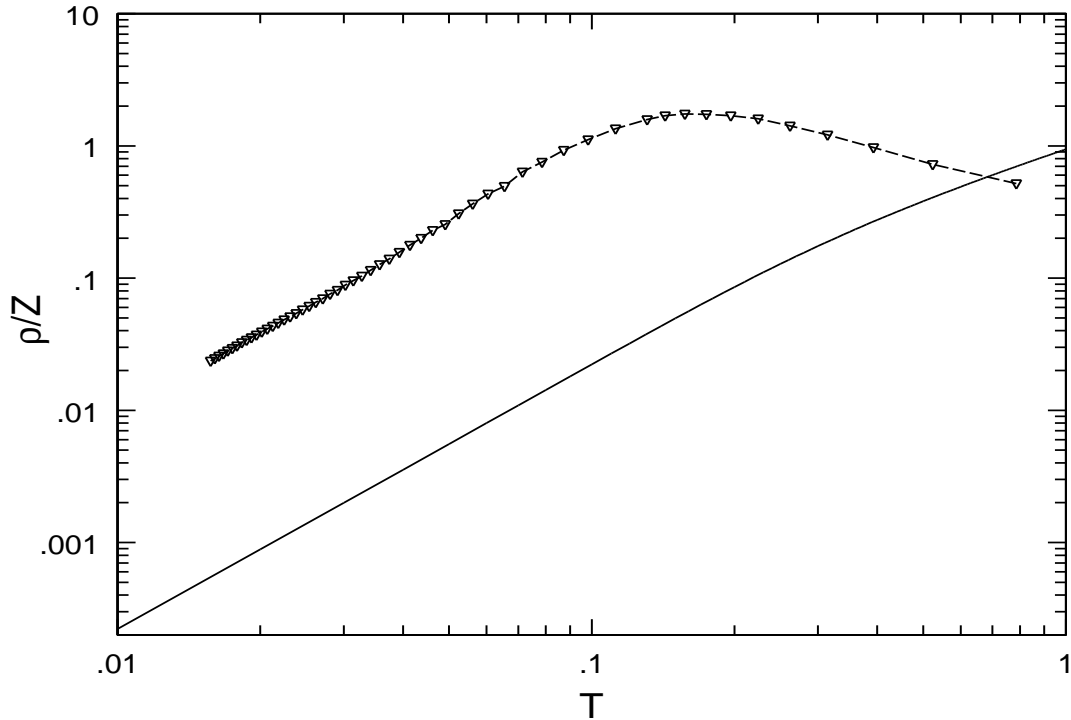

FIG. 4. Same results as in Fig. 3 on double logarithmic scale; same units as in Fig. 3. The $T^2$-behavior at low $T$ is clearly seen.

At low temperatures, $T \lesssim 0.1$, both models show Fermi liquid behavior, i.e. $\rho(T)/Z = AT^2$, as is clearly seen on a double-logarithmic plot (Fig. 4) [35]. The typical $T^2$-dependence originates in the low-frequency behavior of $N_\Sigma(\omega) = -\pi^{-1}\mathrm{Im}\Sigma_R(\omega) =: \widetilde{N}_\Sigma(0)(T^2 + b\omega^2)$ in (21), (22) where $b$ is some constant. Since $(-\partial f/\partial \omega)$ is a strongly peaked function of $\omega$ at the Fermi level, with width $\sim T$, one obtains $\rho(T)/Z = (Z\sigma_{dc})^{-1} \propto \widetilde{N}_\Sigma(0)T^2$ i.e. $A \propto \widetilde{N}_\Sigma(0)$. For spinless fermions with $1/Z$-terms one has $\widetilde{N}_\Sigma(0) \propto U^2/Z$ for all $U$. For the Hubbard model $\widetilde{N}_\Sigma(0)$ can become much larger (this is so even at small $U$, where $\widetilde{N}_\Sigma(0) \propto U^2$ in this case), in particular close to the metal-insulator transition. This explains why $A$ is much larger for the Hubbard model (see Figs. 3,4).

The existence of a maximum in the resistivity for the Hubbard model is in striking contrast to the monotonic increase of $\rho(T)/Z$ in the case of spinless fermions with $1/Z$-terms. This maximum is located almost exactly at the temperature where the system would undergo a phase transition to the antiferromagnetic phase [11] if that ordering were not suppressed. Hence the increase in the resistivity can be attributed to the enhancement in the scattering of electrons by local spin fluctuations. Such an effect does not occur in the case of spinless fermions with $1/Z$-terms: the interaction $U/Z$ is by a factor $1/Z$ weaker and the charge fluctuations are averaged over the very large number of nearest neighbors, i.e. do not act coherently.

At high temperatures the resistivity increases linearly with temperature in both models (for the Hubbard model that part is not shown in Figs. 3,4). This is due to the fact that for $T \to \infty$ the functions $N(\omega), N_c(\omega)$ and $N_\Sigma(\omega)$ become temperature independent, while $(-\partial f/\partial \omega) \to 1/(4T)$, thus leading to $Z\sigma_{dc} \propto 1/T$, i.e. $\rho(T)/Z \propto T$.

## IV. DISCUSSION

We will now discuss the conditions under which the dc-conductivity can be finite at $T > 0$, and to what extent this is a generic property of correlated electron systems in dimensions $d = 1, 2, 3$.

### A. Umklapp scattering

In the absence of phonons and impurities only Umklapp-scattering processes are able to degrade the total crystal momentum. Under such conditions a finite dc-conductivity at $T > 0$ can therefore only occur in a lattice system. In the $d = \infty$ dynamic mean-field theory for the Hubbard model and spinless fermions with $1/Z$-corrections, the correlation problem reduces to a self-consistent single-site problem [22–24], where the importance of the lattice and the notion of reciprocal lattice vectors are no longer directly visible. Nevertheless it must be borne in mind that for quantum mechanical particles the $d \to \infty$ limit can *only* be formulated on a lattice (the lattice constant provides a natural short wavelength cut-off [15]). Hence the results obtained for correlated elecrons in $d = \infty$, in particular those obtained here for the Drude weight $D$ and $\sigma_{dc}$, require the existence of a lattice.

In fact, in $d = \infty$ *all* scattering processes involve Umklapp processes. This is a direct consequence of the topology of the Fermi surface of a lattice system which in the limit $d \to \infty$ is very close to a sphere from which major parts are chopped off at the Brillouin-zone (BZ) boundaries [36]. In fact, the Fermi body covers the fraction $n/g$ of the $(d-1)$-dimensional boundary surface of the hypercubic BZ, where $n$ is the particle density and $g$ denotes the degeneracy of the particle state, i.e. $g = 1$ and $2$ for spinless fermions and the Hubbard model, respectively. Since in $d = \infty$ the Fermi body intersects with the BZ boundary for *all* fillings $n$ it is possible to have Umklapp processes *with arbitrarily small $\vec{q}$-vectors*. There is then no minimum energy $\Delta$ required for Umklapp scattering and hence no exponential suppression $\exp(-\Delta/T)$ of such processes at low temperature. Thus their contribution is always important.

For each dimension $d$ of a hypercubic lattice one may define a critical density $n_{c,d}$ such that for $n \geq n_c^{(d)}$ the Fermi body intesects with the BZ boundaries. While $n_{c,\infty} = 0$ (note, however, that the limits $n \to 0$ and $d \to \infty$ do not commute), one has $0 < n_{c,d} < 1$ for any finite dimension $d$ with $3 \leq d < \infty$, i.e. in general the critical density increases with decreasing $d$. In particular, in $d = 2$ one has $n_{c,2} = 1$, in which case the Fermi body touches the BZ at discrete points, while in $d = 1$ an intersection does not occur for any filling. Hence Umklapp processes become more effective the higher the dimension of the lattice is. Consequently the $d = \infty$ mean-field theory overestimates the effect of Umklapp processes. A finite dc-conductivity is nevertheless a natural feature of lattice systems in sufficiently high dimensions $d$.

### B. Vertex corrections

In the limit $d = \infty$ the dynamic conductivity $\sigma(\omega, T)$ is strictly given by the dressed bubble-diagram since vertex corrections are suppressed [29]. The dc-conductivity $\sigma_{dc}(T)$ is thus determined by the single-particle life time. In general the life-time (or rather its inverse, the scattering rate) receives contributions from scattering processes of quasi-particles involving Umklapp scattering as well as momentum-conserving "normal" scattering. In $d = \infty$ essentially all scattering events involve Umklapp processes. Genuine two-particle interference effects are described by vertex corrections. They enter as soon as $1/Z$-corrections are taken into account and lead, for example, to the renormalization factor $\gamma = 1 + \mathcal{O}(\frac{1}{Z}) > 1$, (18), in the case of spinless fermions. By including $1/Z$ corrections, i.e. vertex corrections, one departs from the $d = \infty$ limit. Thereby the importance of the current-degrading Umklapp processes is reduced. Hence we expect $1/Z$-corrections to *increase* the conductivity.

It is still an unresolved question whether or not vertex corrections are able to restore momentum conservation to such an extent that $\sigma_{dc}(T)$ can be infinite at $T > 0$, i. e. whether $\sigma(\omega, T)$ has a contribution $D(T)\delta(\omega)$ even for $T > 0$. If at all, this would be a property of systems where Umklapp processes are relatively unimportant, such as *low*-dimensional system away from half-filling [9]. Recently Castella et al. [13] conjectured that a divergent dc-conductivity at $T > 0$ is a generic property of *integrable* system such as the Luttinger model, interacting bosons and the $U = \infty$ Hubbard model in $d = 1$. In this case our finding of a finite dc-conductivity at $T > 0$ would be generic in any dimension since most interaction models appear to be non-integrable.

It will be very interesting to investigate further whether, and if so under what conditions, the dc-conductivity of correlated lattice electrons may diverge at $T > 0$ even in dimensions $d = 2$ or 3.

We are grateful to H. Fukuyama for generating our interest in the problem, to G. Czycholl, W. Metzner, E. Müller-Hartmann for valuable discussions, and to W. Hanke, P. Horsch, S. Maekawa and Th. Pruschke for useful information. This work was supported in part by the SFB 341 of the Deutsche Forschungsgemeinschaft.